\documentclass[12pt,amstex,aps,amsfonts,prsty]{article} 
\usepackage[dvips]{graphicx}
\usepackage{amssymb}
\usepackage{amsmath}
\usepackage{a4wide}
\usepackage{citesort}

\begin{document}

\begin{titlepage}

\begin{center}

{\LARGE \bf Monolayer Spreading on a Chemically Heterogeneous Substrate}

\vspace{0.3in}

{\Large \bf N. Pesheva$^{1}$ and G. Oshanin$^2$}

\vspace{0.1in}

{\large \sl $^1$ Institute of Mechanics, Bulgarian Academy of Sciences,\\ 
Acad. G. Bonchev St. 4, 1113 Sofia, Bulgaria
}

{\large \sl $^2$ Laboratoire de Physique Th{\'e}orique des Liquides, \\
Universit{\'e} Paris 6, 4 Place Jussieu, 75252 Paris, France
}

\begin{abstract}
We study the spreading kinetics of a monolayer of hard-core particles 
on a semi-infinite, chemically heterogeneous solid substrate, 
 one side of which is coupled to a particle reservoir.
The substrate is modeled as a square lattice containing two types of sites  --
ordinary ones and special, chemically active  sites  
placed at random positions with
 mean concentration $\alpha$. These special sites temporarily immobilize 
 particles of the monolayer which then serve as impenetrable 
obstacles for the other particles. 
In terms of a mean-field-type theory, we show that the mean displacement 
$X_0(t)$ of the monolayer edge grows with time $t$ 
as $X_0(t) = \sqrt{2 D_{\alpha} t \ln(4 D_{\alpha} t/\pi a^2)}$,
($a$ being the lattice spacing). 
This time dependence is confirmed by numerical simulations; $D_{\alpha}$ is obtained numerically 
for a wide range of
values of the parameter $\alpha$ and trapping times of the chemically active sites.
 We also study numerically the behavior of a stationary particle current in finite samples.
The question of the influence of attractive particle-particle interactions 
on the spreading kinetics is also addressed.
\end{abstract}

\end{center}

\vspace{0.2in}


\end{titlepage}

\section{Introduction}

The stability and spreading kinetics of ultrathin wetting
 films on solid substrates 
are of technological and scientific importance in 
many applications
ranging from coatings, paints, dielectric layers, 
thin film lubrication, microelectronic devices, to 
fundamental studies of adsorption and particle dynamics  
\cite{1,cazabat,granick,leger}. 
In the case of homogeneous, chemically pure 
substrates, the properties of such films 
are relatively well understood
through a series of experimental and theoretical works 
\cite{1,cazabat,granick,leger,2,90,91}.

However, most of the naturally occurring surfaces used 
in thin film experiments are chemically 
heterogeneous on nanometer to micrometer scales, e.g. 
due to contamination, cavities, uneven oxide layer, etc. 
On the other hand, deliberately tailored chemically heterogeneous 
substrates are also increasingly being used for engineering of 
desired nano- and micropatterns in thin films (see, e.g. Refs. \cite{31,32}). 
In addition, some recent studies 
have revealed a possibility of controlling the growth of biological systems 
by attaching them to structured surfaces \cite{33}
and to recognize biological molecules, (e.g., proteins), selectively by 
bringing them into contact with nanostructured surfaces \cite{34}.  

A considerable amount of recent 
theoretical, numerical and experimental work 
has been devoted to the analysis of $\it equilibrium$  properties of thin
films on chemically heterogeneous substrates. These studies focused mostly 
on such issues  as stability of films, pattern formation, 
appearance of self-organized structures, as well as
the impact of chemical disorder on 
the contact angle and appropriate generalization of the Young's equation 
\cite{16,17,18,19,20,21,22,23}. Much less is known, however, on
spreading kinetics of 
ultrathin liquid films on chemically disordered surfaces. 
Here, the only available studies 
concern Molecular Dynamics simulations \cite{adao1,adao2} and experimental
 analysis \cite{prl,lan} 
of precursor films spreading on substrates with chemically impure sites. 
To the best of our knowledge,
no theoretical analysis has been as yet performed.

In the present paper, motivated by recent experimental studies of precursor 
films spreading on chemically disordered substrates \cite{prl,lan},
we analyse the spreading kinetics of molecularly thin films 
on substrates with 
randomly placed chemically active sites.
We focus here on
systems with the so-called planar geometry, i.e. on
systems in which film's thickness (or concentration of particles in the film
in case of monolayers) varies effectively only  along one spatial
coordinate. This typical experimental situation occurs when a solid, which
may be  a plane or a cylindrical fiber, is immersed in a
liquid bath. Here, the particle concentration in the liquid film, which
extracts from the macroscopic meniscus and climbs along the solid, varies
only with the altitude above the edge of the macroscopic meniscus and is
independent  of the perpendicular, horizontal coordinate. 
The meniscus then serves as a reservoir of particles,
which is in equilibrium with the spreading monolayer and "feeds" it. 
The solid substrate is modeled here in a usual fashion as a 
regular, square lattice of adsorption sites; 
chemical heterogeneity is introduced by adding some concentration $\alpha$ of 
special, chemically active sites, which
temporarily trap moving particles which then become obstacles for others. We analyse
 here, both analytically and numerically, the behavior of the mean displacement $X_0(t)$ 
of the monolayer edge. In terms of a mean-field-type approach, we find that 
$X_0(t)$ grows with time $t$ as
 $X_0(t) = \sqrt{2 D_{\alpha} t \ln(4 D_{\alpha} t/\pi a^2)}$,
($a$ being the lattice spacing). 
This time dependence, which contains a non-trivial logarithmic factor, 
is confirmed by the numerical simulations. 
As well, we obtain  $D_{\alpha}$ numerically for a wide range of values of chemical sites' concentration
 $\alpha$ and of the
trapping times. 
 We also consider the situation when our substrate is of a finite extent
along the $X$-axis and study numerically the behavior of the stationary particle current.
The question of the influence of attractive particle-particle interactions 
on spreading kinetics is also addressed.

The paper is structured as follows: In Section 2 we formulate a microscopic stochastic
model of spreading kinetics. In Section 3 we 
 derive basic equations and present their mean-field-type 
solution appropriate for situations with annealed spatial distribution 
of the chemically active sites.
In Section 4 we 
describe our Monte Carlo simulations model.
Results of Monte Carlo simulations of 
spreading kinetics of monolayers composed of hard-core particles
and analysis of the behavior of the particle current in finite samples are 
presented in Section 5. 
Next, in Section 6 we consider spreading behavior in the case when the 
monolayer particles experience 
short-range, nearest-neighbor attractive interactions. 
Finally, in Section 7  we conclude with a summary and discussion of our results.

\section {The Model}

As we have already remarked, our model is
relevant to the following experimental situation. Suppose that a vertical
solid wall is immersed in a bath of liquid. The liquid interface, which is
initially horizontal, changes its shape in the vicinity of the solid wall
and a macroscopic meniscus builds up.  The size of the
macroscopic meniscus (both horizontally and vertically) is  comparable to
the capillary length. After a suitable transient period an ultrathin liquid film 
(a monolayer)
exudes from the static macroscopic meniscus and climbs up the solid wall \cite{cazabat}.
In Ref. \cite{90} a microscopic stochastic model describing
 spreading kinetics of molecular films on chemically 
homogeneous, ideal substrates has been developed. Here we extend this approach on the 
situation when chemical disorder is present.

Particles dynamics on the solid surface is generally regarded as  an
activated random hopping motion, constrained by hard-core interactions,
between the local minima of a wafer-like array of potential wells. Such
wells occur because the monolayer's particles
experience short-range forces exerted by the atoms of the solid.
Consequently, the interwell distance $a$ is related to the
spacing between the atoms of the substrate. Without going into details of
the particle-substrate interactions, we suppose that for
the transition to one of the neighboring potential wells a particle has to
overcome a potential barrier. This barrier does not create a preferential
hopping direction, but results in a finite time interval $\tau $ between the
consecutive hops, defined through the Arrhenius formula. 

To specify the positions of the wells, we introduce a pair of perpendicular
coordinate axes ($X,Y$), where $X$ is a vertical coordinate, which measures
the altitude of a given well above the meniscus (a reservoir), 
while $Y$ defines the
horizontal position of this well. For simplicity, we suppose that the lattice of 
potential wells is a regular square lattice of spacing $a$ (see Fig. 1). It will be made clear 
below that the effects we observe do not drastically depend on the precise form of the
underlying lattice. 

Further on, we assume that the substrate contains some concentration $\alpha$ of immobile, 
chemically 
active sites, placed at random positions. For simplicity, we suppose that the 
spatial distribution of 
these sites is commensurate with
the underlying lattice of potential wells, such 
that the chemically active sites can be viewed as 
occupying random positions 
on the sites of the square lattice exactly (see, Fig. 1). 

We turn next to the definition of the hopping probabilities. We suppose that the 
latter are symmetric regardless of whether a particle occupies an 
ordinary or a chemically active site.
In the former case, a particle chooses a jump direction with the same
probability equal to $1/4$, which means that
being on ordinary site a particle always attempts to perform a hop.
On the contrary, in the latter case, there is a probability that 
the particle stays at the site -- a pausing probability $\epsilon$, 
which mirrors the chemical specificity of sites and
 hence, results in a temporal trapping effect.
Here, the particle
selects the jump direction with probability 
$(1 - \epsilon)/4$, where the parameter $\epsilon$ can be expressed
as 
\begin{equation}
\epsilon = 1 - \exp\Big(U_{tr}/k_B T\Big),
\end{equation}
$k_B T$ being
the temperature measured in the units of the Boltzmann constant $k_B$, 
while $U_{tr}$ denotes  the trapping energy, $U_{tr} < 0$. 
Note that the typical time $\tau^*$ spent by a given 
particle being trapped by a chemically active site is just $\tau^* = \tau/(1-\epsilon)$.

Consequently, the site-dependent jump direction probabilities $p(X,Y)$ can be written down as
\begin{equation}
p(X,Y) = \left\{\begin{array}{ll}
1/4,     \mbox{ if the site $(X,Y)$ is an ordinary site,} \nonumber\\
(1-\epsilon)/4,     \mbox{  if the site $(X,Y)$  is a chemically active site.}
\end{array}
\right.
\end{equation} 
After the jump direction is chosen, the particle attempts to hop onto the target site. The 
jump is fulfilled if the target site is empty at this moment of time; otherwise, the particle remains at its position.  

Finally, we view
the liquid bath 
as a reservoir of particles (of an infinite
capacity) which maintains a constant concentration 
$C_0$ of fluid particles
at the edge of the macroscopic meniscus, i.e. the line $X = 0$ in Fig. 1 
(see Ref. \cite{90} for more details).
Here, for simplicity, we take $C_0 = 1$. 
The behavior for arbitrary $C_0$ will be considered elsewhere \cite{nina}. 

We hasten to remark that dynamics in
 disordered lattice gas-type models, relevant to the one employed here,  
has been extensively studied within different contexts, including, for instance, 
charge carrier transport in dynamic percolating systems \cite{yossi},
tracer diffusion within the first layers of solid surfaces \cite{olivier1} and in adsorbed monolayers \cite{olivier}, 
 tracer and collective diffusion on solid surfaces \cite{kreuzer,gomer}, in pure 
and disordered crystals \cite{binder,kehr} or 
collective diffusion in zeolites \cite{3,4,5}. 
The systems analyzed in these works differ, however, 
considerably from the situation under study; 
here we present a first, to the best of our knowledge,
 lattice gas-type description of $spreading$ dynamics of 
monolayers on substrates with chemical disorder.

To close this section  
it might be instructive to discuss the limitations 
of such a non-interacting 
lattice-gas-type model. In the "real world" systems, 
the particles appearing on top of a solid substrate -- adsorbed particles, experience two 
types of interactions: namely, 
interactions with the atoms of the underlying solid -- the solid-particle (SP) interactions, 
and mutual interactions with each other -- the particle-particle (PP) interactions. 
The SP interactions  are characterized by a repulsion at short scales,  
and an attraction at longer distances. The repulsion
 keeps the adsorbed particles 
some distance apart of the solid, while attraction
favors adsorption and   
hinders particles desorption as well as migration along the solid surface. 
In this regard, our model corresponds to the regime of the so-called
intermediate localized adsorption \cite{1,200}: the particles forming a
 monolayer are neither completely fixed
in the potential wells created by the SP 
interactions, nor completely mobile.
This means, the potential wells are rather deep with 
respect to the particles desorption (desorption barrier $U_d \gg k_B T$), so that only an
adsorbed monolayer can exist, but have a much lower energy barrier $V_l$ against 
the lateral movement across the
surface,  $U_d \gg V_l > k_B T$. 
In this regime, any monolayer 
particle spends a considerable part of its time at the bottom
of a potential well and jumps sometimes, solely due to the thermal activation, 
from one potential minimum to another
in its neighborhood; after the jump is performed, the particle dissipates all its energy to the host solid. 
Thus, on a macroscopic time scale 
the particles do not possess any velocity. 
The time $\tau$ separating two
successive jump events, is just 
the typical  time a given particle 
spends in a given well 
vibrating around its minimum; as we have already remarked,  $\tau$ 
is related to the temperature, the barrier for the
lateral motion and the frequency of the
solid atoms' vibrations by the  Arrhenius formula.

We emphasize that such a type of random 
motion is essentially different from the standard hydrodynamic picture of particles
random motion in the two-dimensional "bulk" liquid phase, e.g. in 
free-standing liquid films, 
in which case there is a
velocity distribution and spatially $random$ motion results from the PP scattering. 
In this case,
the dynamics can be only approximately considered as an activated hopping of particles, confined to some effective
cells by the potential field of their neighbors, along a lattice-like structure of such cells (see, e.g.
Refs. \cite{300,400}). In contrast to the dynamical model to be studied here, standard two-dimensional hydrodynamics
presumes that the particles do not interact with the underlying solid. In 
realistic systems, of course, both the
particle-particle scattering and scattering by the potential wells due to the interactions with the host solid, (as
well as the corresponding dissipation channels), are  important \cite{kreuzer,600}. 
In particular, it has been shown that addition of dissipation to the host solid removes
the infrared divergencies in the dynamic density correlation 
functions and thus makes the transport coefficients finite 
\cite{700}. 
On the other hand, homogeneous adsorbed monolayers may only exist
 in systems in which the
attractive part of the PP
interaction potential is essentially  weaker than that
 describing interactions with the solid; otherwise, such
monolayers become unstable and dewet spontaneously from the solid surface.
As a matter of fact, for stable homogeneous monolayers,  
the PP interactions are at least ten times weaker that the
interactions with the solid atoms \cite{200}. 

Consequently, the standard hydrodynamic picture of 
particles dynamics is inappropriate 
under the defined above physical conditions. Contrary to that, 
any adsorbed particle moves due to random hopping events, 
activated by chaotic vibrations of the solid
atoms, along the local minima of an array of potential wells, 
created due to the interactions with the solid
\cite{1,200}. As we have already remarked, in the physical conditions 
under which such a dynamics takes place, 
the PP interactions are much weaker than the SP interactions
and hence do not perturb significantly
the regular array of potential wells due to the SP interactions. In our model, 
we discard completely the attractive part of
the PP interaction potential and take into account only the repulsive one, 
which is approximated by an abrupt, hard-core-type
potential.   

The question of the monolayer spreading in the 
case when some short-range attractive particle-particle interactions are 
present will be briefly addressed in Section 6.

\section{Basic equations and a mean-field-type solution}

Let $\rho_t(X,Y)$ denote the local density of the monolayer particles 
at time moment $t$ at the site $(X,Y)$. 
This local density obeys the following balance equation
\begin{eqnarray}
\label{q}
\tau \frac{d \rho_t(X,Y)}{d t} &=& -  p(X,Y) \;
\rho_t(X,Y) \sum_{(X',Y')} \Big(1 - \rho_t(X',Y')\Big) + \nonumber\\
&+& \Big( 1 - \rho(X,Y)\Big) \sum_{(X',Y')} p(X',Y') \rho_t(X',Y'),
\end{eqnarray}
where $(X',Y')$ denotes a nearest-neighboring to $(X,Y)$ site, while the summation symbol
 with the subscript 
 $(X',Y')$ means that the summation extends over all nearest to   $(X,Y)$ sites. Note that 
the factors $ \Big(1 - \rho_t(X',Y')\Big)$ and  $\Big( 1 - \rho(X,Y)\Big)$ on the right-hand-side of
Eq. (\ref{q}) account 
for the steric constraints due to hard-core interactions and represent the (decoupled) probabilities
that the target sites are unoccupied at time moment $t$.

Equation (\ref{q}) holds for all particles except for the rightmost particles for each fixed $Y$, since 
for the latter, by definition, 
the hops away of the monolayer (i.e. such that increase their $X$ position to $X + a$) 
are not constrained by the hard-core interactions. Let now $X_0(Y,t)$ denote the 
$X$-position of the rightmost particle in the column
 with fixed $Y$. Evidently, one has for
  $\rho_t(X=X_0(Y,t),Y)$ the following equation
\begin{eqnarray}
\label{qq}
\tau \frac{d \rho_t(X_0(Y,t),Y)}{d t} &=& - p(X_0(Y,t),Y) \;
\rho_t(X_0(Y,t),Y) \sum_{Y'=Y \pm a} 
 \Big(1 - \rho_t(X_0(Y,t),Y')\Big) - \nonumber\\
&-&  p(X_0(Y,t),Y)  \; \rho_t(X_0(Y,t),Y) \;\Big(1 - \rho_t(X_0(Y,t)-a,Y)\Big) - \nonumber\\
&-&  p(X_0(Y,t),Y) \; \rho_t(X_0(Y,t),Y) + \nonumber\\
&+& \Big( 1 - \rho(X_0(Y,t),Y)\Big) \Big[ \sum_{Y'=Y \pm a}   p(X_0(Y,t),Y')  \rho_t(X_0(Y,t),Y') + \nonumber\\
&+&  p(X_0(Y,t)+a,Y) \; \rho_t(X_0(Y,t) + a,Y) \Big] + p(X_0(Y,t)-a,Y),
\end{eqnarray}
which thus has a different structure compared to Eq. (\ref{q}). 
Note that the last term on the right-hand-side of Eq. (4)
is not multiplied by neither the occupation factor 
$\rho_t(X_0(Y,t),Y)$ nor by the steric factor 
$(1 - \rho_t(X_0(Y,t),Y))$. This happens, namely, 
because the last term describes the event in which the 
rightmost particle, present, by definition, at the site $X_0(t) - a$, (i.e. $\rho_t(X_0(Y,t) - a,Y) = 1$),
hops at the vacant site $X_0(t)$, (i.e. $\rho_t(X_0(Y,t),Y) = 0$).

We turn next to the mean-field-type picture 
assuming first that chemically active sites are uniformly spread along 
the substrate with mean density $\alpha$, and $p(X,Y)$ is a position-independent constant
\begin{equation}
p(X,Y) \approx \frac{p_{\alpha}}{4}
\end{equation}
An estimate of $p_{\alpha}$ will be presented below.
                                    
Then,  we note that
the dependences of $\rho_t(X,Y)$ on the $X$ and the $Y$ coordinates have 
quite different origins. There is a reservoir of particles, which maintains
fixed occupation of all sites at $X = 0$. 
Consequently, we may expect a
regular $X$-dependence of $\rho_t(X,Y)$. In contrast, the $Y$-dependence may
be only noise; the uniform boundary at the $X = 0$ 
insures that there is no
regular dependence on the $Y$ coordinate and, in absence of disorder in the jump direction probabilities, 
only the particle dynamics may
cause fluctuations in $\rho_t(X,Y)$ along the $Y$-axis. Hence, following Ref. \cite{90}
we will disregard these fluctuations  and suppose that the local density
varies along the $X$-axis only, i.e. $\rho_t(X,Y) = \rho_t(X)$.
Consequently, we will have an effectively  one-dimensional  problem in which
the presence of the $Y$-direction will be accounted only through the
particles' dynamics. We note finally that assumption of such a type is, in
fact, quite consistent with experimental observations \cite{cazabat},
which show that in case of sufficiently smooth substrates and liquids  with
low volatility the width of the film's front is very narrow.

Then, in neglect of the fluctuations along the $Y$-axis the variable $\rho_t(X)$ 
can be viewed as a local time-dependent variable describing
occupation of the site $X$ in a $stochastic$ $process$ in which hard-core
particles perform hopping motion (with a time interval $\tau^* $ between the
consecutive hops) on a one-dimensional lattice of spacing $a$ connected, at the site $X = 0$, to
 a particle reservoir which maintains constant occupation of this site.

For $t \gg \tau$, characteristics of such a process are then described by the 
following nonlinear system of coupled equations.
The mean displacement of the rightmost particle (the monolayer edge) obeys:
\begin{equation}
\label{m1}
\tau \; \frac{d X_{0}(t)}{d t} =   \frac{a p_{\alpha}}{4} \; \rho_t(X=X_0(t)),
\end{equation}
where $\rho_t(X)$ is determined by
\begin{equation}
\label{k}
\tau \; \frac{\partial \rho_t(X)}{\partial t} = \frac{a^2  p_{\alpha}}{4} \; \frac{\partial^2 \rho_t(X)}{\partial X^2},
\end{equation}
which holds for $0 \leq X \leq X_0(t)$ and is to be solved subject to two boundary conditions:
\begin{equation}
\label{n}
\rho_t(X = 0) = 1,
\end{equation}
and 
\begin{equation}
\label{m2}
a \; \tau \; \frac{\partial \rho_t(X = X_0(t))}{ \partial t} = - \left. \frac{a^2  p_{\alpha}}{4} \;  
 \frac{\partial \rho_t(X)}{\partial X}\right|
_{(X=X_0(t))} \; - \; \tau \; \rho_t(X= X_0(t))  \frac{d X_0(t)}{d t}
\end{equation}
These two boundary conditions mimic, first, the presence of a particle reservoir, and second, 
show that for the rightmost particles of the monolayer the jumps away of the monolayer are 
not constrained by hard-core interactions.

We note now that Eqs. (\ref{m1}) to (\ref{m2}) constitute a classical mathematical 
problem of solving a partial differential equation
with one of the  boundaries being 
imposed in the moving frame, 
which is akin to the so-called Stefan problem. Its solution can be 
found in a standard way by 
observing that the density profiles $\rho_t(X)$
written in terms of the scaling variable $\omega = X/X_0(t)$ become stationary. In the limit 
$t \gg \tau$, the mean displacement of the monolayer edge thus follows
\begin{equation}
\label{mu}
X_0(t) = \sqrt{2 D_{\alpha} t \ln\Big(\frac{4 D_{\alpha} t}{\pi a^2}\Big)},
\end{equation}
where $D_{\alpha}$ is given by
\begin{equation}
\label{d}
D_{\alpha} = \frac{a^2  p_{\alpha}}{4 \tau}.
\end{equation}
In a similar fashion, one finds that the total number M(t) 
of particles,
\begin{equation}
M(t) = \int^{\infty}_0 dX \rho_t(X),
\end{equation}
emerged on the substrate up to time $t$, obeys
\begin{equation}
M(t) \sim \sqrt{\frac{4 D_{\alpha} t}{\pi}},
\end{equation}
which implies that the mean density in the monolayer slowly decreases with time
\begin{equation}
\overline{\rho_t(X)} = \frac{M(t)}{X_0(t)} \sim \sqrt{\frac{2}{\pi \ln\Big(4 D_{\alpha} t/\pi a^2\Big)}}
\end{equation}
Note that dependence of $\overline{\rho_t(X)}$ on disorder, which enters only through
the effective diffusion coefficient $D_{\alpha}$ is logarithmically weak. Note also
that the mean displacement $X_0(t)$ of the monolayer edge
grows at a faster rate than the conventionally expected pure diffusive
 $\sqrt{t}$-law due 
to an additional factor $\sqrt{\ln(t)}$; consequently, fitting of experimental 
curves or numerical results with a  pure 
$\sqrt{t}$-law is meaningless since the effective diffusion coefficient will 
appear to depend on time of observation. 
In Fig. 2 we present numerical evidence 
of this additional logarithmic factor. 
In Fig. 3 we depict numerical results describing the behavior of $D_{\alpha}$.
Analytical estimates of $D_{\alpha}$ will be presented elsewhere \cite{nina}.

We finally remark that within the employed 
mean-field dynamical approach, we can also obtain 
an average stationary particles current $<J_{part}>$. 
Solving Eq. (\ref{k}) subject to the reservoir boundary condition in Eq. (\ref{n}), 
as well as imposing a trapping boundary condition at the right edge of the 
substrate, $\rho_t(X = N) = 0$,
we find that
\begin{equation}
<J_{part}> = \frac{D_{J}}{N},
\end{equation}  
i.e. the current has a Fickian dependence on the substrate's length.

The effective diffusion coefficient $D_J$ can be estimated within a mean-field-type approximation
as follows: 
in the stationary state it matters actually 
how much time, on average, a given particle spends on a given  lattice site. Such an average time is, evidently,
\begin{equation}
\label{tau}
<\tau> = \tau \times (1 - \alpha) + \tau^* \times \alpha,
\end{equation}  
where the first term represents a contribution of ordinary sites, while the second one gives
an average time spent by a given particle on chemically active sites. 
Consequently, the effective diffusion coefficient $D_J$ 
can be estimated as
\begin{equation}
\label{DJ}
D_J = \frac{a^2}{4 <\tau>} = 
\frac{a^2}{4 \tau}\frac{1 - \epsilon}{1 - \epsilon (1 - \alpha)}
\end{equation}
This result is, of course, exact for $\alpha = 0$ and $\alpha = 1$, i.e. 
for chemically homogeneous substrates. It appears that it describes 
reasonably well (see Fig. 4) the numerical data for 
$\alpha \sim 1$ and arbitrary $\epsilon$, as well 
for small values of $\epsilon$ and arbitrary $\alpha$.

\section{Numerical simulations}

In our simulation algorithm, we follow closely  the model defined in Section 2.
We consider a square lattice 
$\Lambda$  with linear sizes $L_x$ and $L_y$  and with every lattice site 
$(X,Y)$ we associate an occupation variable $n_{(X,Y)}$  which may assume  only
 two values $\{+1,0\}$. The value $+1$ signifies that the site $(X,Y)$ is occupied, while
$0$ means that this site is vacant.   

The initial  configuration is an empty lattice except  for  the zeroth raw ($X = 0$).
The left edge of the system $X = 0$ is coupled to a 
particle reservoir which keeps the zeroth  raw
always occupied ($C_0=1$),
 i.e. $\{n_{(0,Y)} \equiv 1, Y=1,\dots ,L_y\}$.
In the $Y$-direction periodic boundary conditions are imposed to reduce the finite-size effects. 
The right edge of the system
is coupled to an empty reservoir, so that the raw $\{X=L_x+1\}$ is always empty. 
Note that such a formulation
allows us to study both dynamic and static characteristics. While studying spreading dynamics, we take 
$L_x$ sufficiently large and take care that displacement of the rightmost particle in each 
column $Y$ is less than $L_x+1$.
When studying the behavior of the stationary particle current, we focus on $L_x$ 
not that large and let the system evolve until the density profiles in the system attain
a stationary state.

The following time-saving procedure has been implemented.
At every non-normalized time step $i$ a particle in the system is chosen at random. Let the particle's
coordinates be denoted by $(X,Y)$.
Then the particle may either stay at the site $(X,Y)$ with probability $\epsilon(X,Y)\{=\epsilon, 0\}$, or with
an equal probability, $p(X,Y)={(1-\epsilon(X,Y)) / 4}$, 
may attempt to jump onto 
one of the neighboring sites, chosen at random. The jump is actually fulfilled if the target site is
empty. Otherwise,  the particle remains at the site $(X,Y)$.
If the initial site is in the zeroth raw and if after the update the particle
moves it is  immediately filled by a particle from the reservoir and 
the number of particles $N_{i}$ in the system
is increased. If the initial site is in the last raw $X=L_x$ and 
if after the update the particle moves to $X'=L_x+1$ the number of particles in the system
is decreased.
The time is renormalized according to
   \begin{eqnarray}
  t_{i+1}   =t_i + {1\over N_{i+1}}\, ,
   \end{eqnarray}
where $N_{i+1}$ is the total number of particles in the system at the non-normalized time $(i+1)$.
We use the averaged renormalized time in our studies of the time-dependent quantities.
 
Most of the simulations are performed for a system of size $100\times 25$, 
$200\times 50$ and $100\times 100$
 in units of  the lattice constant $a$. 
Larger system sizes are also  considered in few cases.  
The results are usually averaged over $N_s=2,5,10$ different substrates and for 
each substrate $N_r=5,10$ different runs are performed. Typical Monte Carlo
simulation lasted $1.6  \div  2\times 10^5$ MCS per site.

\section {Simulation Results} 

After passing through a transient regime the system reaches a stationary 
non-equilibrium state characterized by a stationary average particle current 
 $J_{part}$ flowing through the system and a constant
average density gradient.

We studied here how do both,
the spreading diffusion coefficient $D_{\alpha}$ and the diffusion 
coefficient in the
stationary state $D_J$, depend on the pausing probability 
$\epsilon$ and on the concentration $\alpha$
of the chemically active  sites.
The spreading diffusion coefficient $D_{\alpha}$ was 
determined from the time dependence of the
average interface position $X_0(t)$ before particles start leaving the right edge of the system.
It appears that the law in Eq. (\ref{mu}) describes very well the time behavior of the average 
interface position
not only for $\alpha=0, \epsilon=0$ \cite{90}, 
but also for practically the  whole interval of values
of $\alpha$ and $\epsilon$, except at $\epsilon=1.0$, i.e. infinitely deep trapping sites (see Fig. 2).

For determination of the diffusion coefficient $D_J$ in the
stationary state we use Fick's law, $J_{part}=-D_J \nabla \rho$,  
 by measuring the average particle
current, $J_{part}$ (per site), and the average density gradient,
 $\nabla \rho(\approx const.)$,
 in the stationary state at given pausing probability
$\epsilon$ and concentration $\alpha$.
The obtained results for the spreading diffusion coefficient 
$D_{\alpha}$ are presented in Fig. 3
and the corresponding results for the
diffusion coefficient $D_J$ in the stationary state are given in Fig. 4.
Curiously enough, the values found for
 $D_{\alpha}$ are always lower then those obtained for $D_J$.

We turn now to the special case when the pausing probability on chemically active sites
is $\epsilon=1$. The specific feature of this case is that the particle, once arriving at any 
chemically active site stays there forever, serving then as impenetrable obstacle for the other particles. 
It means that in this case one has an induced $percolative$ behavior.  
The time behavior of the average interface position 
for $\alpha> 0.1$ is no longer fitted well 
by the function in Eq. (\ref{mu}) (one expects that here a logarithmic time behavior should take place)
and the above mentioned method cannot
 be employed  to determine the spreading diffusion coefficient $D_{\alpha}$.
For given $\alpha$ the averaged density distribution in the stationary state is still constant
and $\nabla \rho\approx - (1-\alpha)/L_x$.  In order to get a reliable estimates for the
studied  quantities (e.g. the particle current $J_{part}$) 
the demand on the computing time as the concentration $\alpha \to \alpha_c$
increases significantly since longer
time runs are necessary as well as averaging over more substrates is needed
and finally also  bigger systems should be simulated.
The approximate value found for the concentration $\alpha_c \approx 0.4 \pm 0.01$ at which the 
particle current $J_{part}$ (respectively $D_J$) turns to zero is consistent
with $1-p_c$, where $p_c=0.592746$ \cite{ZS87} is the critical probability for
site percolation in the square lattice (see Fig. 5).

\section {Monolayer of interacting particles.}

We turn finally to the case when the monolayer particles experience short-range (nearest-neighbor) attractive 
interactions.
Let us consider 
the simplest possible case when the corresponding Hamiltonian is
   \begin{eqnarray}
   H=-U\sum_{(X',Y')}  n(X,Y) n(X',Y') \ ,
   \end{eqnarray}
where $U\; (U>0)$ is the constant describing the attraction between two diffusing particles and
the summation symbol with the subscript "(X',Y')" 
means that summation extends over 
the sites $(X',Y')$, neighbouring to the site $(X,Y)$.
 We still  assume the activation mechanism for the 
hopping motion of the monolayer particles;
that is, the probability for jump depends
on the trapping energy of the site $(X,Y)$ through:
$$P_{jump}(X,Y)=\exp\left({U_{tr}(X,Y) \over k_BT}\right)
$$

We take into consideration the interaction between the diffusing particles by assuming that
the particle "feels" the other particles when choosing the direction for the jump, i.e.:
$$
P_{dir}((X,Y),(X',Y'))={1\over Z} \exp \left({-\Delta H((X,Y),(X',Y')) \over 2K_BT}\right) \ , \  
Z=\sum_{(X',Y')}  P_{dir}((X,Y),(X',Y')) \ ,
$$
where 
$$
 \Delta H((X,Y),(X',Y'))  =  H(X',Y')-H(X,Y) 
$$
and $H(X,Y)  =  -U \; n(X,Y) \sum_{(X',Y')} n(X',Y')$
is the interaction energy of the particle at the site $(X,Y)$.

For high enough temperatures one may, 
as a first approximation, try to determine
the diffusion coefficients $D_{\alpha}$ and $D_J$ in the same way 
as it was done for the non-interacting system.
The temperature dependence of the diffusion coefficients determined in this way
is shown in Fig. 6.
 As could be seen taking into consideration the interaction between
 the diffusing particles leads to a decrease of the
 diffusion coefficients. For higher temperatures the effect
is less pronounced.

For lower temperatures another method for 
determining  $D_J$ should be employed.
While for lower temperatures  the time behavior of the average interface position 
$X_0(t)$ is still reasonably well  described 
by Eq. (\ref{mu}), the density distribution along the spreading direction  in the
stationary state is no longer linear.  
In  Fig. 7  the corresponding density distributions  
(for homogeneous substrate, $\alpha =0$, $U_{tr}=0$)
are shown for three different temperatures for the interacting system in 
the stationary state. One can see that
at $k_BT=0.5U$ there is clearly a phase separation though there is a stationary 
particle current flowing through the system. The interface between the two phases
is approximately at $X=L_x/2$. 
At higher temperatures, e.g.  $k_BT=2.5U$ the density distribution is getting closer
to a linear distribution (as in the non-interacting case) but is still  not linear.
This system is very similar to the driven diffusive system introduced by Katz, 
Lebowitz and Spohn
\cite{KLS83} where there is a stationary particle current flowing in the system
due to a bias in the transition rates.

\section{Conclusions}

To conclude, we have studied 
the spreading kinetics of a monolayer of hard-core particles 
on a semi-infinite, chemically heterogeneous solid substrate, 
one side of which is attached to a reservoir of particles.
The substrate is modeled as a square lattice containing two types of sites - 
ordinary ones and special, chemically actives sites  
placed at random positions with
 mean concentration $\alpha$. These special sites temporarily immobilize 
the particles of the monolayer which then serve as impenetrable 
obstacles  for the other particles. 
In terms of a mean-field-type theory, we have shown that the mean displacement 
$X_0(t)$ of the monolayer edge grows with time $t$ 
as $X_0(t) = \sqrt{2 D_{\alpha} t \ln(4 D_{\alpha} t/\pi a^2)}$,
($a$ being the lattice spacing).
This  nontrivial time
dependence is confirmed by the numerical simulations. 
 For a broad range of values of $\alpha$ and of the
trapping times of the chemically active sites 
(pausing probabilities) $D_{\alpha}$ has been obtained from extensive Monte Carlo simulations.
In addition, we have studied numerically the behavior of the stationary particle current in
 finite samples.
We have observed that, curiously enough, the diffusion coefficient $D_{\alpha}$ deduced from 
the analysis of the data on the spreading kinetics, and the 
one obtained from the analysis of the data on the stationary particle currents, 
$D_J$, are different from each other
and obey $D_{\alpha} < D_J$. 
Besides, we have 
found that the system displays a percolation-type behavior when
$\epsilon = 1$ and $\alpha \to \alpha_c  \approx 0.4 \pm 0.01$. 
In this limiting case
 both $D_{\alpha}$ and $D_J$ vanish.
The question of the influence of attractive particle-particle interactions 
on spreading kinetics has been also addressed.  We have observed that taking 
into consideration attractive interactions between
 the diffusing particles leads to a decrease of the
 diffusion coefficients. For higher temperatures the effect
becomes less pronounced, as it should. Finally, we have found that 
for sufficiently strong attractions 
the density distribution along the spreading direction 
 in the
stationary state is no longer linear and  that
there is clearly a phase separation, though the stationary 
particle current does not vanish.

\vspace{0.5cm}
\begin{center} {\bf Acknowledgments}
\end{center}

The authors thank M.Vou\'e and J.De Coninck 
for fruitful discussions and interest to this work. We also wish 
to thank M.Vou\'e for his 
help in preparing the preliminary simulation codes.

\newpage

\section*{Figure captions}

\begin{itemize}

\item {\bf Fig. 1.} Schematic representation of the monolayer 
in contact with a particle reservoir on a chemically heterogeneous substrate. 
Gray squares denote chemically active sites.

\item {\bf Fig. 2.} Plot of $X_0(t)/\sqrt{t}$ versus $\sqrt{ln(t)}$  - 
numerical evidence for time-dependent 
logarithmic corrections to the mean displacement of the monolayer edge. Circles
denote the time moment 
when the rightmost particles of the monolayer reach the right edge of the substrate, 
such that the finite-size effects come into play.

\item {\bf Fig. 3.} The dependence of the spreading 
diffusion coefficient $D_{\alpha}$ is shown:
(a) as a function of the pausing probability
 $\epsilon$ at different fixed concentrations $\alpha$ of the chemically active sites,
solid circles-solid line   -- $\alpha=0.1$, solid up-triangles-solid  line  -- $\alpha=0.3$,  solid
squares-solid  line  -- $\alpha=0.5$, solid diamonds-solid  line  -- $\alpha=0.7$, 
solid down-triangles-solid   line  -- $\alpha=0.9$;
(b) as a function of the concentration $\alpha$ at different fixed pausing probabilities $\epsilon$,
solid circles-solid  line  -- $\epsilon=0.1$, solid up-triangles-solid  line  -- $\epsilon=0.3$,  solid
squares-solid  line  -- $\epsilon=0.5$, solid diamonds-solid  line   -- $\epsilon=0.7$, 
solid down-triangles-solid  line -- $\epsilon=0.9$.

\item {\bf Fig. 4.} 
The dependence of the  diffusion coefficient in the stationary state $D_J$ is shown:
(a) as a function of the  pausing probability $\epsilon$ at different fixed concentrations $\alpha$
of the chemically active sites,
solid circles-solid  line -- $\alpha=0.1$, solid up-triangles-solid line   -- $\alpha=0.3$,  solid
squares-solid  line  -- $\alpha=0.5$, solid diamonds-solid  line   -- $\alpha=0.7$, 
solid down-triangles-solid  line  -- $\alpha=0.9$;
(b) as a function of the concentration $\alpha$ at different fixed pausing probabilities $\epsilon$,
solid circles-solid   line   -- $\epsilon=0.1$, solid up-triangles-solid  line  -- $\epsilon=0.3$,  solid
squares-solid   line -- $\epsilon=0.5$, solid diamonds-solid   line  -- $\epsilon=0.7$, 
solid down-triangles-solid line  -- $\epsilon=0.9$.
The dotted lines are the corresponding analytical curves given by Eq. 17.

\item {\bf Fig. 5.} Percolation threshold. The plot of $D_J$ versus $\alpha$ for $\epsilon=1$. Linear 
extrapolation of the numerical data gives the critical value of $\alpha = \alpha_c$ at which the current 
vanishes equal to $\alpha_c \approx  0.4 \pm 0.01$.

\item {\bf Fig. 6.} 
The temperature dependence of the spreading diffusion coefficient
$D_{\alpha}$ and the diffusion coefficient $D_J$ in the stationary state.  Concentration of the chemically active sites is
$\alpha=0.5$ of the trapping sites  and their trapping energy $U_{tr}$ is taken equal to $U_{tr}=-0.7$.
For the non-interacting system $(U=0)$ the results for $D_J$ and for $D_{\alpha}$ are given 
by solid squares-solid  lines and by  open squares-dotted line, respectively.
For the weakly (compared to the trapping energy)
 interacting system $(U=0.1)$ --  solid up-triangles-solid  line 
and  open up-triangles-dotted  line depict, respectively, the behavior of  $D_J$ and
$D_{\alpha}$. For $(U=0.3)$ -- solid circles-solid  line  define $D_J$, while open circles-dashed line
determine the corresponding behavior of the spreading diffusion coefficient $D_{\alpha}$.

\item {\bf Fig. 7.} 
The average density distributions along the spreading direction $X$ in homogeneous systems 
($\alpha =0$ and $U_{tr}=0$) are  shown for the interacting $(U=1)$ system 
$100\times 25$ (in units of  the lattice constant)
 in  the stationary state at  
three different temperatures:
 $k_BT=2.5 U$ solid  line, $k_BT=1 U$ dashed line, 
$k_BT=0.5 U$ dotted line. 

\end{itemize} 

\pagebreak

\begin{figure}[ht]
\begin{center}
\includegraphics*[scale=0.9]{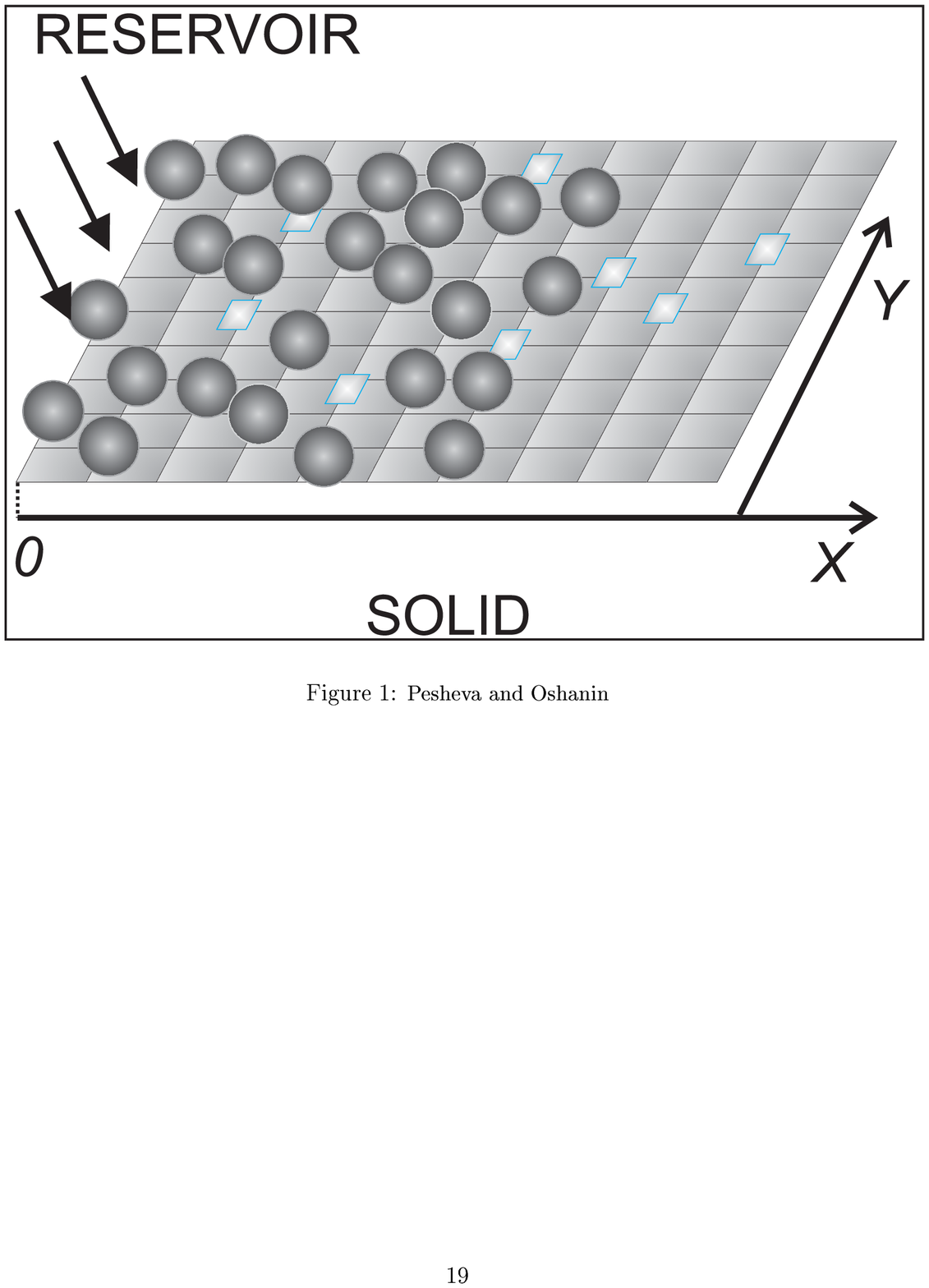}
\caption{\label{NF1} {\small Pesheva and Oshanin}}
\end{center}
\end{figure}

\pagebreak

\begin{figure}[ht]
\begin{center}
\includegraphics*[scale=0.8]{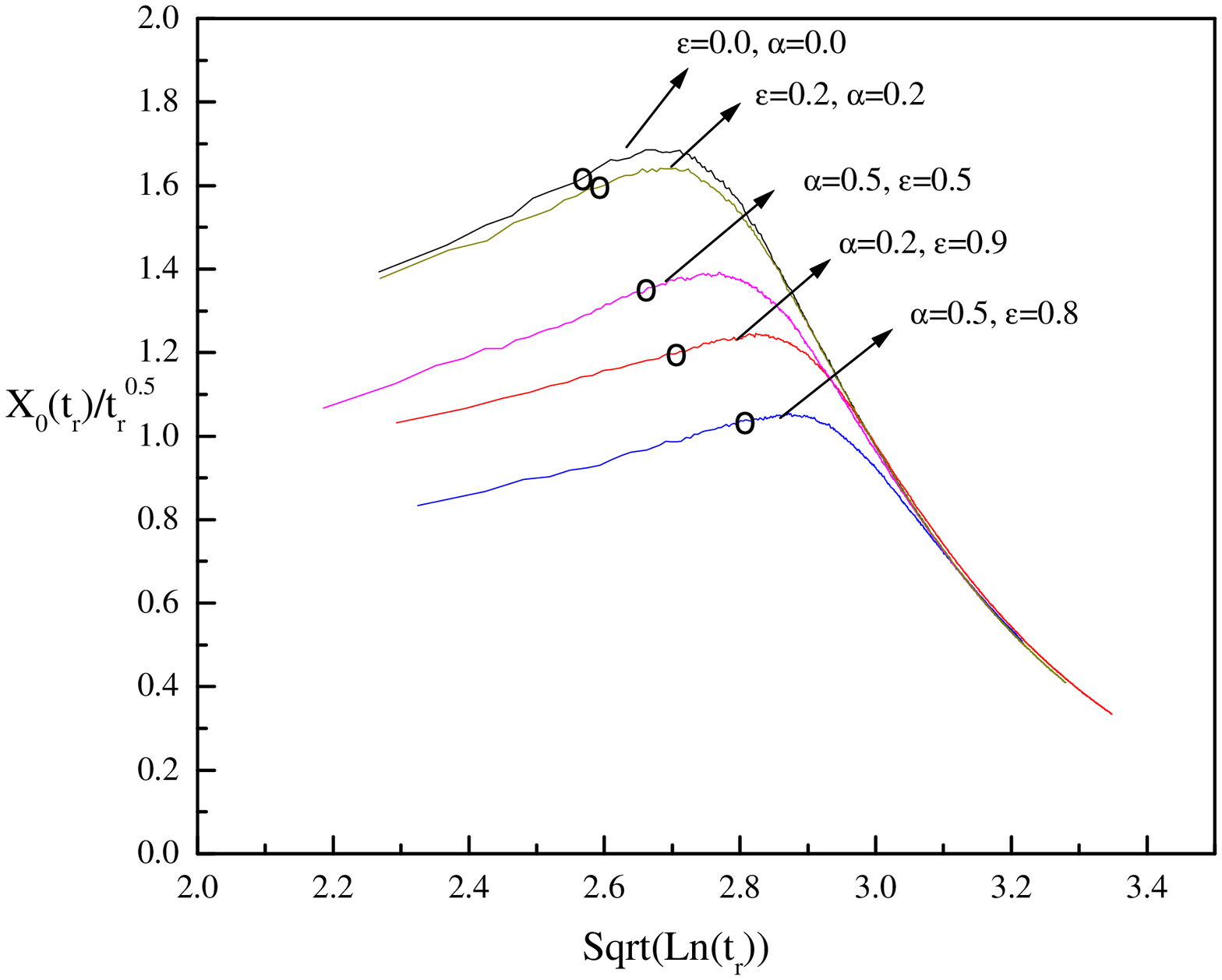}
\caption{\label{NF2} {\small Pesheva and Oshanin}}
\end{center}
\end{figure}

\pagebreak

\begin{figure}[ht]
\begin{center}
\includegraphics*[scale=0.7]{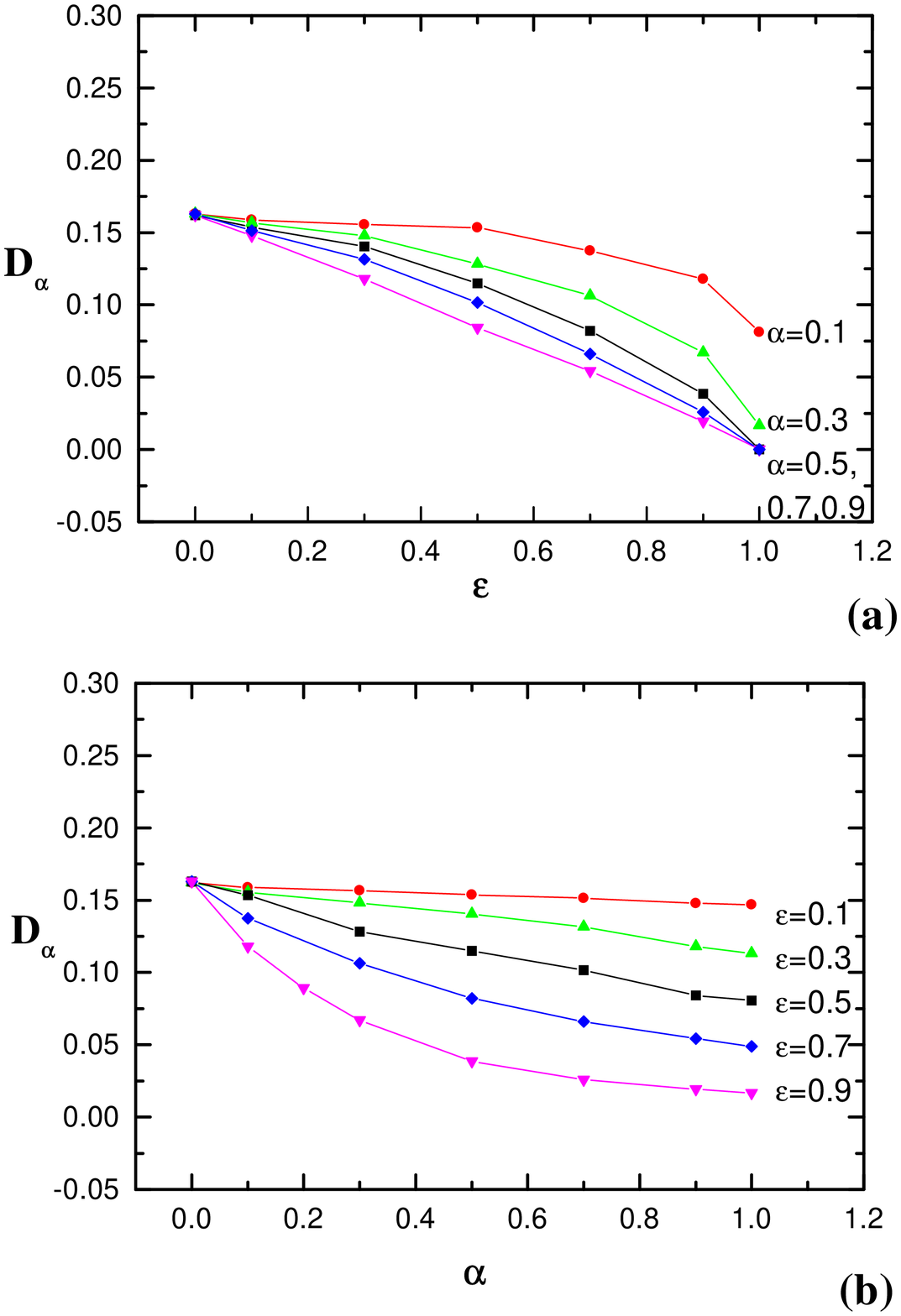}
\caption{\label{NF3} {\small  Pesheva and Oshanin}}
\end{center}
\end{figure}

\pagebreak

\begin{figure}[ht]
\begin{center}
\includegraphics*[scale=0.7]{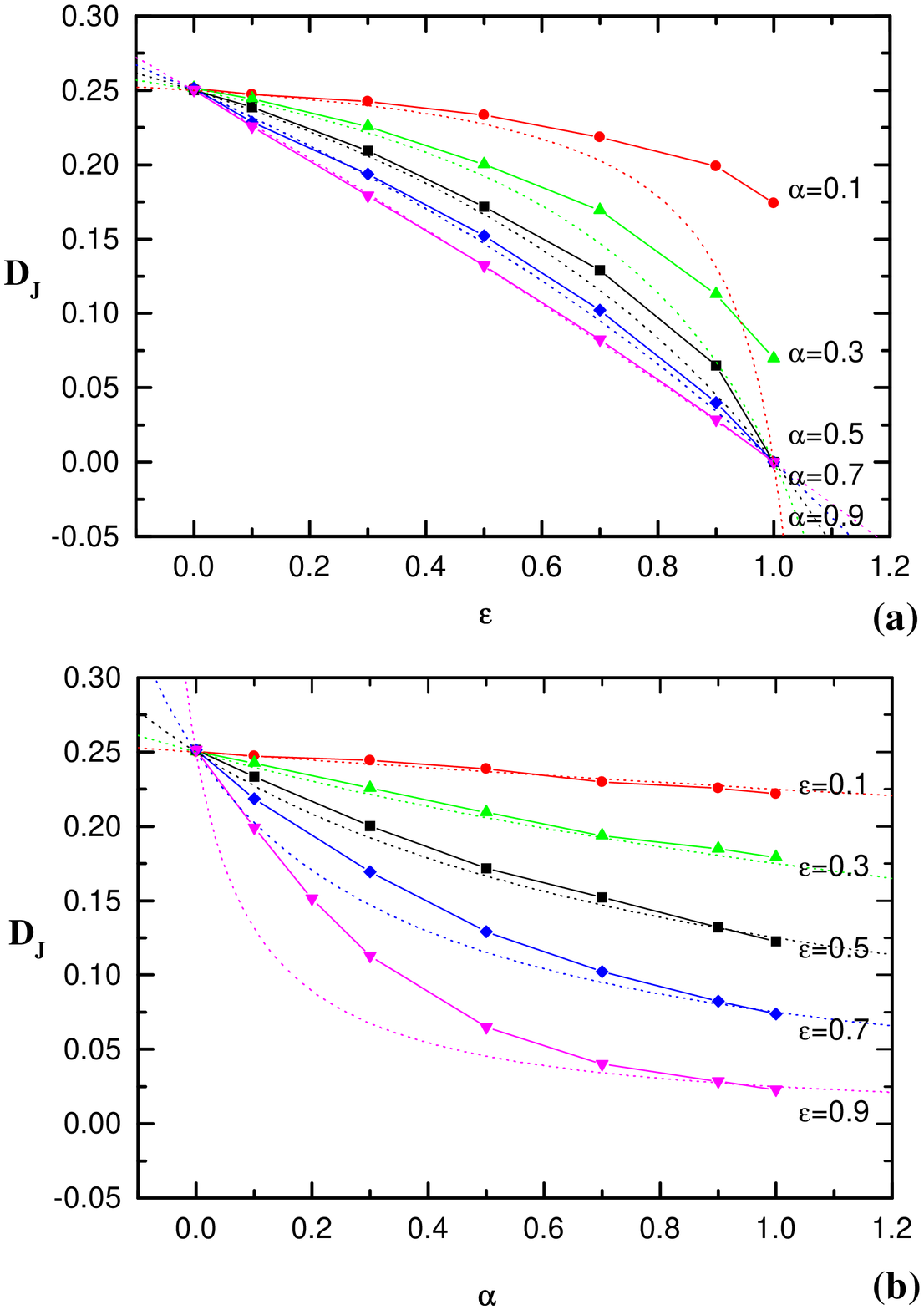}
\caption{\label{NF4} {\small Pesheva and Oshanin}}
\end{center}
\end{figure}

\pagebreak

\begin{figure}[ht]
\begin{center}
\includegraphics*[scale=0.9]{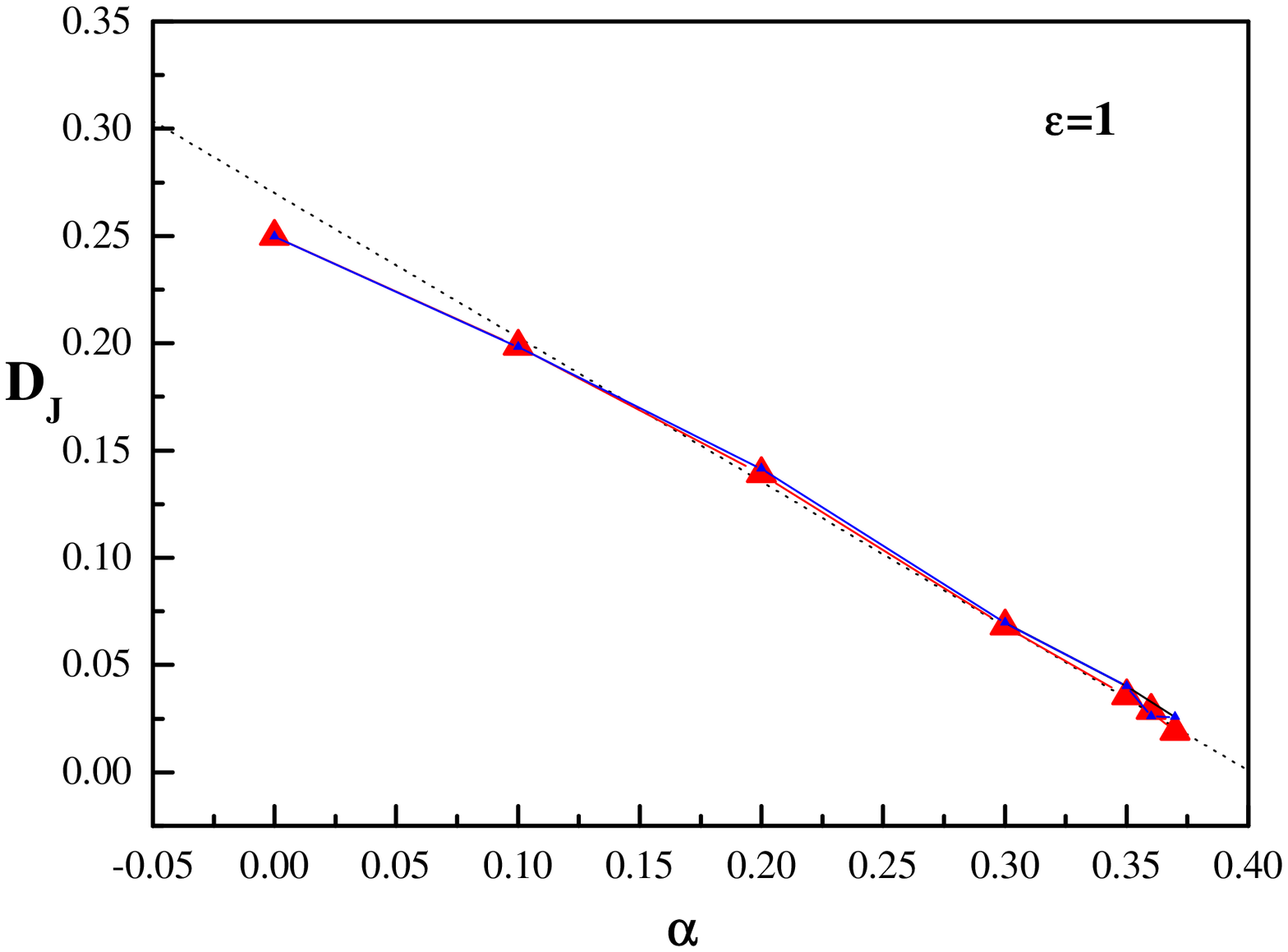}
\caption{\label{NF5} {\small Pesheva and Oshanin}}
\end{center}
\end{figure}

\pagebreak

\begin{figure}[ht]
\begin{center}
\includegraphics*[scale=0.9]{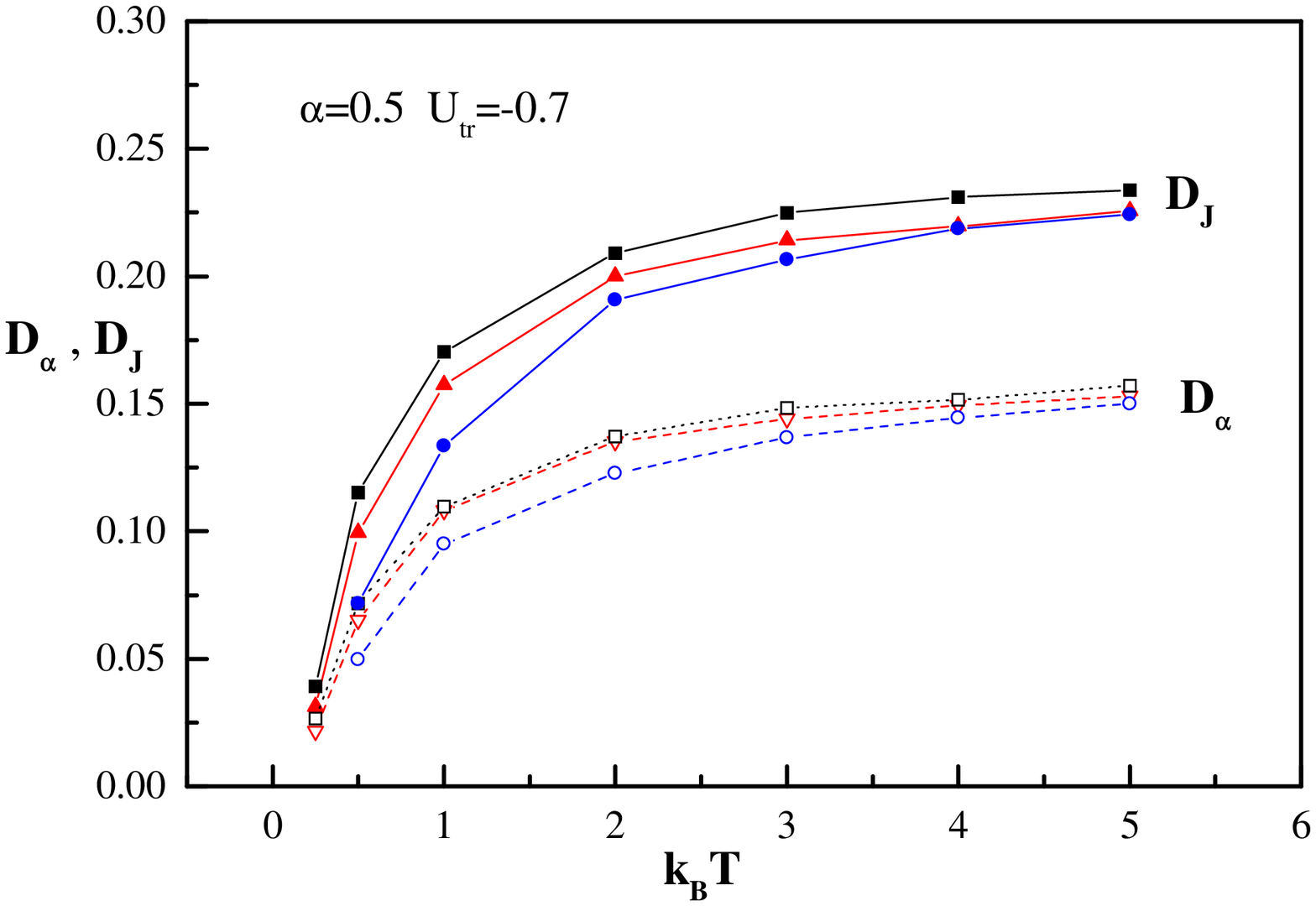}
\caption{\label{NF6} {\small Pesheva and Oshanin}}
\end{center}
\end{figure}

\pagebreak

\begin{figure}[ht]
\begin{center}
\includegraphics*[scale=0.9]{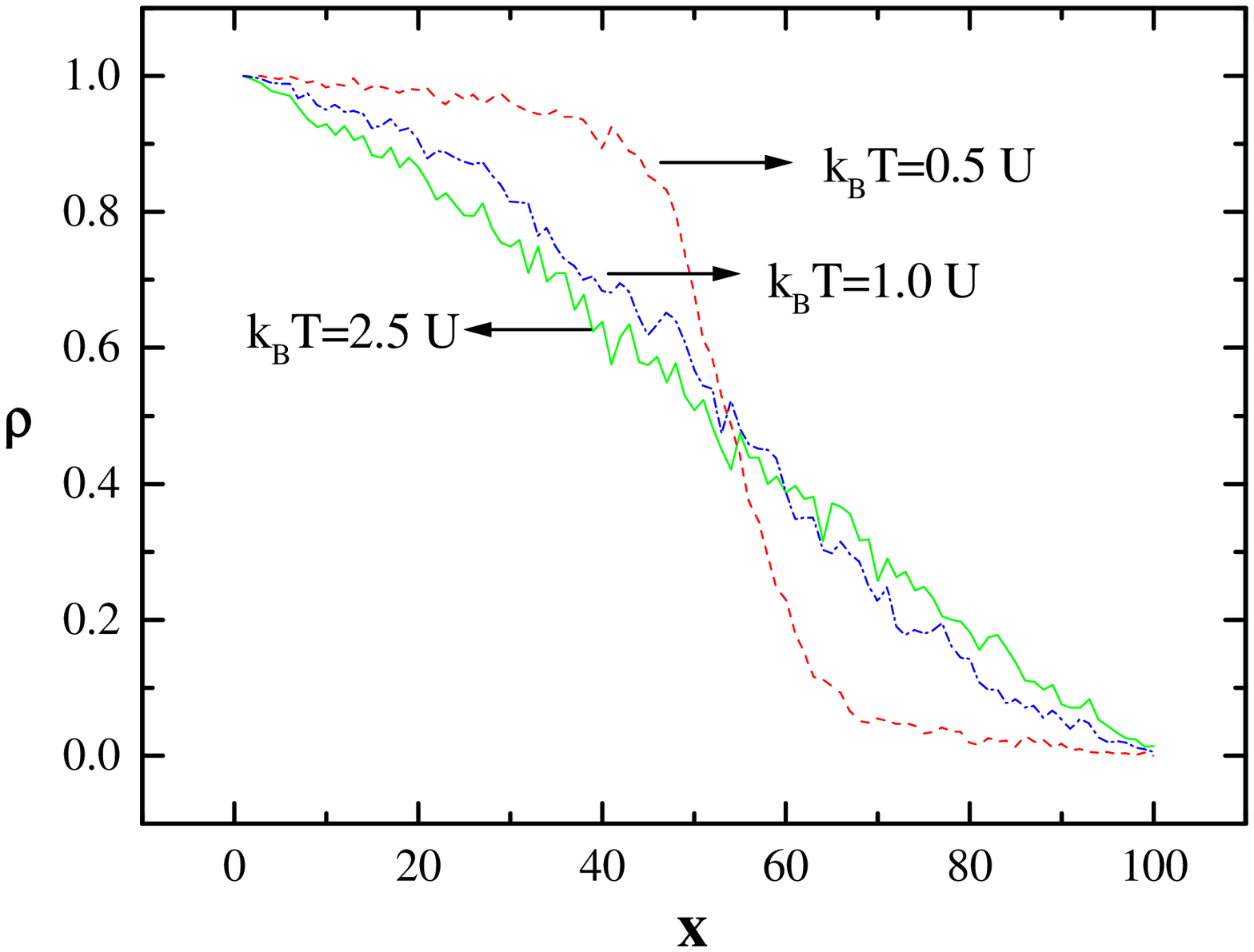}
\caption{\label{NF7} {\small Pesheva and Oshanin}}
\end{center}
\end{figure}

\end{document}